%% file: eprint.tex
\documentclass[12pt]{article}
\usepackage{graphicx}
\usepackage{amsmath}
\usepackage{caption}
\usepackage{subcaption}

\newcommand\pubnumber{}  
\newcommand\pubdate{\today}

\def\institute{$^{1}$Institute of Experimental Physics, SAS, Ko\v{s}ice\\
$^{2}$P.J. \v{S}af\'{a}rik University, Ko\v{s}ice\\}

\def\Title#1{\begin{center} {\Large #1 } \end{center}}
\def\Author#1{\begin{center}{ \sc #1} \end{center}}
\def\Address#1{\begin{center}{ \it #1} \end{center}}

\newcommand\pubblock{\rightline{\begin{tabular}{l} \pubnumber\\
         \pubdate  \end{tabular}}}
\newenvironment{Abstract}{\begin{quotation}  }{\end{quotation}}
\newenvironment{Presented}{\begin{quotation} \begin{center} 
             PRESENTED AT\end{center}\bigskip 
      \begin{center}\begin{large}}{\end{large}\end{center} \end{quotation}}

\input econfmacros.tex

\begin{document}
\begin{titlepage}
\pubblock

\vfill
\Title{Top quark pair asymmetries}
\vfill
\Author{F. Sopkov\'{a}$^{1,2}$} 
\Address{\institute}
\vfill
\begin{Abstract}
In this article, we study top quark pair charge, energy and
  incline asymmetry at parton level using Powheg Monte-Carlo
  generator.
\end{Abstract}
\vfill
\begin{Presented}
$9^{th}$ International Workshop on Top Quark Physics\\
Olomouc, Czech Republic,  September 19--23, 2016
\end{Presented}
\vfill
\end{titlepage}
\def\thefootnote{\fnsymbol{footnote}}
\setcounter{footnote}{0}

\section{Introduction}

\begin{figure}[h]
\centering
\includegraphics[height=1.2in]{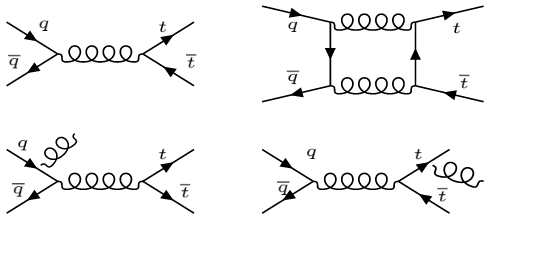}
\caption{The example SM $t\bar{t}$ diagrams resulting in the charge asymmetry at NLO.
}
\label{fig:NLO}
\end{figure}

The charge asymmetry in top quark pair ($t\bar{t}$)
 production causes top and antitop quarks to have different kinematic distributions. In the Standard Model (SM), the effect occurs only at next to leading order (NLO) in quantum chromodynamics. 
The charge asymmetry is present
 in charge asymmetric initial states (quark-antiquark annihilation), the gluon fusion does not contribute to charge asymmetry. 
The examples of the most important $t\bar{t}$ Feynman diagrams in SM are in the Figure~\ref{fig:NLO}. The interference of Born and box diagrams gives a positive contribution to the charge asymmetry, a negative contribution comes from the interference of diagrams containing initial and final state gluon radiation. 
 Using Monte Carlo generator POWHEG/hvq~\cite{POWHEG_manual} 10 million proton-proton collisions with top-antitop pair production in dilepton channel at $\sqrt{s}$ = 13\, TeV  were generated. The composition of initial states are: gluon-gluon initial state (69\%), quark/antiquark-gluon initial state (25\%), quark-antiquark initial state (6\%). Analysis was performed at parton level.

\section{Charge asymmetry}

The charge asymmetry on parton level was estimated using the top, antitop and positive, negative lepton rapidties,
  respectively. The rapidity of a particle along a beam is defined as $y=1/2\ln[(E+p_{z})/(E-p_{z})]$, where $E$ is the energy of particle and $p_{z}$ is longitudinal component of particle momentum. 


The charge asymmetry is defined as

\begin{equation} \label{asym-y}
A_{C}^{t\bar{t}}=\frac{N\left(\Delta|y|>0\right)-N\left(\Delta|y|<0\right)}{N\left(\Delta|y|>0\right)+N\left(\Delta|y|<0\right)}
\end{equation}

where $N(\Delta|y|>0)$ is the number of events with $\Delta|y|>0$
 and  $\Delta|y|=|y_{t}|-|y_{\bar{t}}|$ is defined in laboratory frame \cite{t_charge_asym}.

The asymmetry was studied as a function of $t$$\bar{t}$ invariant mass $m_{t\bar{t}}$, $t\bar{t}$ transverse momentum ($p_{T}^{t\bar{t}}$), $t\bar{t}$ pair longitudinal velocity ($\beta_{z}^{t\bar{t}}$), defined as $\beta={|p_{t}^{z}+p_{\bar{t}}^{z}|}/{(E_{t}+E_{\bar{t}})}$, where $p^{z}$ and $E$ are the longitudinal momentum and energy in laboratory frame \cite{t_charge_asym}. The aim of the analysis is to find optimal range of these kinematic variables, where the relative precision of the asymmetry is highest.

\begin{figure}[!h]
\centerline{
  \begin{subfigure}[b]{0.35\textwidth}
    \includegraphics[width=\textwidth]{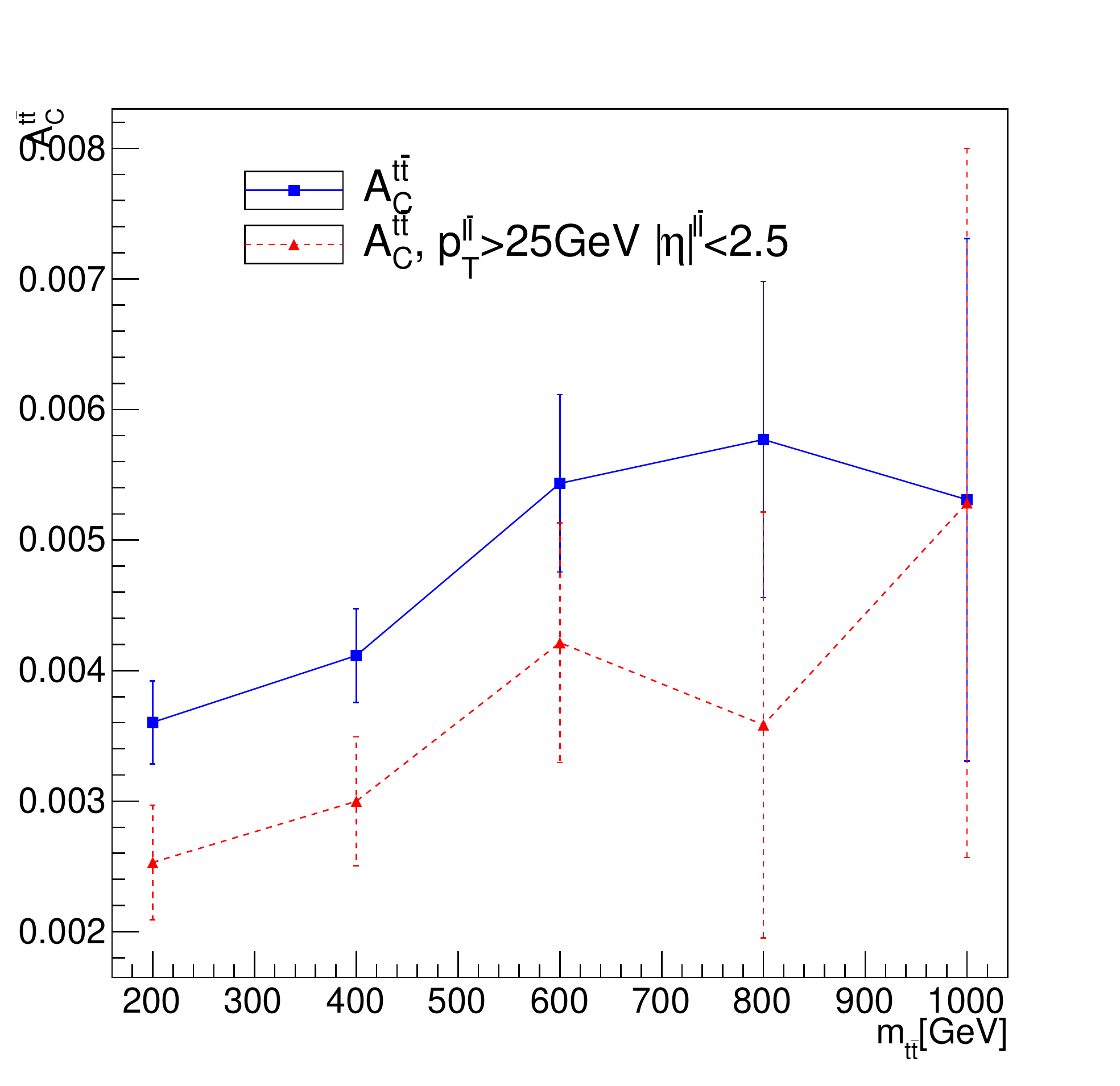}
    \caption{$A_{C}^{t\bar{t}}$ vs $m_{t\bar{t}}$}
    \label{fig:f1}
  \end{subfigure}
  \begin{subfigure}[b]{0.35\textwidth}
    \includegraphics[width=\textwidth]{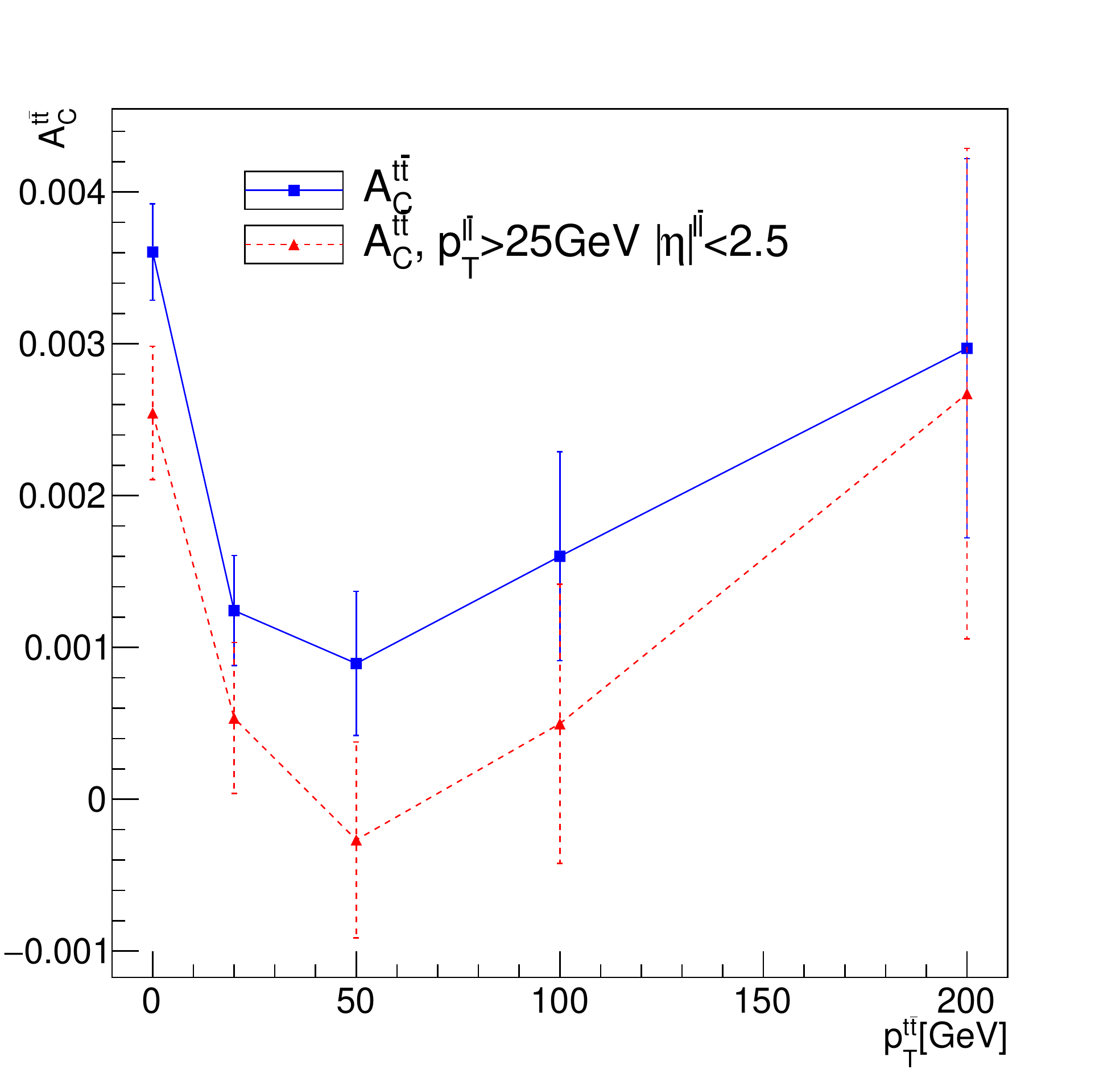}
    \caption{$A_{C}^{t\bar{t}}$ vs $p_{T}^{t\bar{t}}$}
    \label{fig:f2}
  \end{subfigure}
  \begin{subfigure}[b]{0.35\textwidth}
    \includegraphics[width=\textwidth]{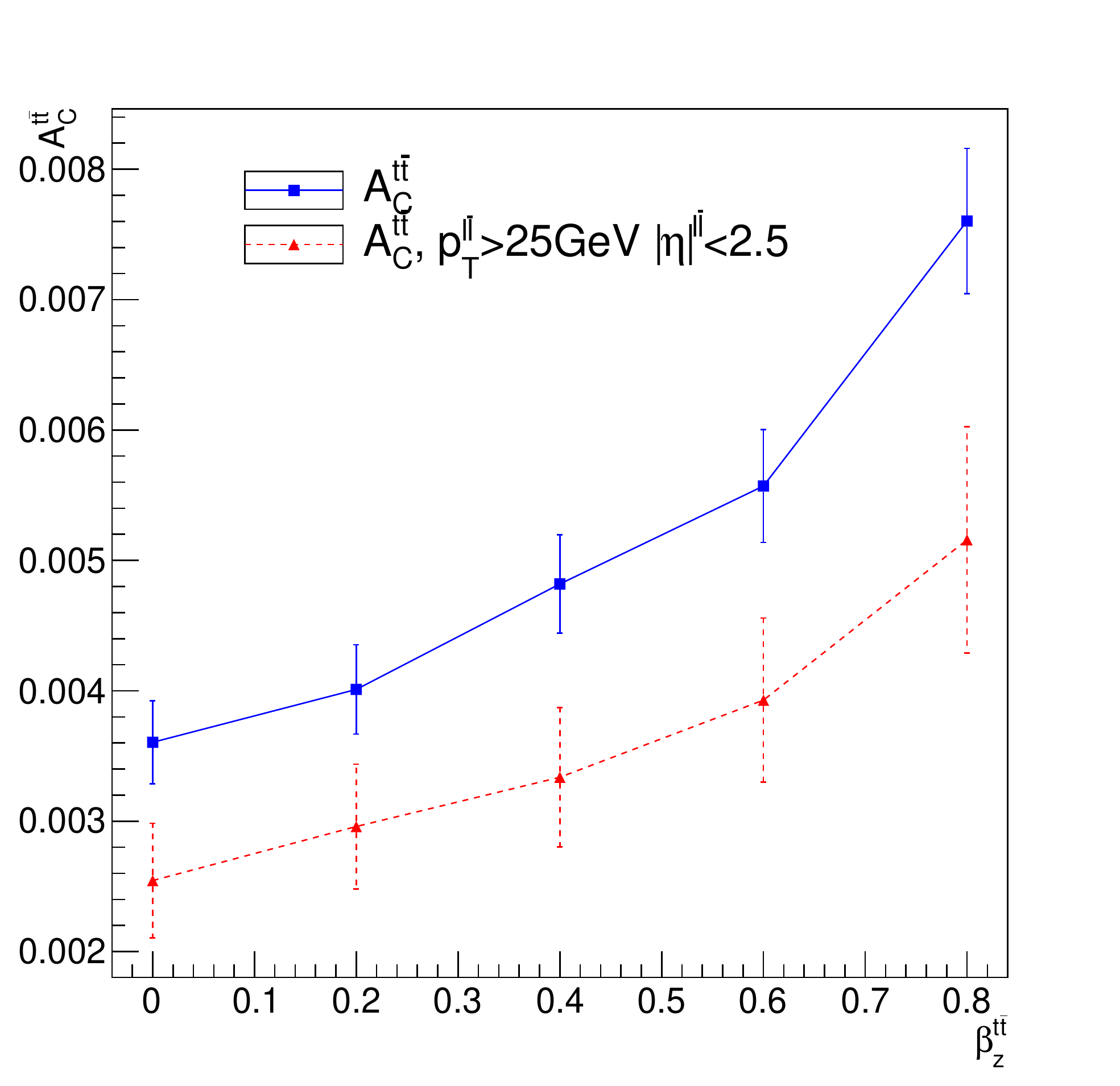}
    \caption{$A_{C}^{t\bar{t}}$ vs $\beta_{z}^{t\bar{t}}$}
    \label{fig:f3}
  \end{subfigure}}
  \caption{The $t\bar{t}$ integral charge asymmetry
 as a function of cut on $m_{t\bar{t}}$(left), $p_{T}^{t\bar{t}}$(middle),
$\beta_{z}^{t\bar{t}}$(right).  }
  \label{fig:Detector-cut-for-tt}
\end{figure}

In Figure~\ref{fig:Detector-cut-for-tt}, the integral asymmetry,
i.e. asymmetry as a function of a cut on a given kinematic variable
 (e.g. for $m_{t\bar{t}}$ dependence, this means $A_{C}^{t\bar{t}} = A_{C}^{t\bar{t}}(m_{t\bar{t}} > cut)$) is presented.
Without applying any cuts,  the asymmetry shows increasing trend as a function of $m_{t\bar{t}}$ and $\beta$ and shows the largest asymmetry at very lowest $p_{T}^{t\bar{t}}$ as it is expected. To come closer to the experimental conditions the following cuts are applied for both leptons transverse momenta and pseudorapidity $p_{T}^{\ell} >$25\,GeV, $|\eta|^{\ell}<$2.5. After applying such cuts on leptons, the
asymmetries are smaller, see Figure~\ref{fig:Detector-cut-for-tt}.



\begin{figure}[!h]
\centerline{
  \begin{subfigure}[b]{0.35\textwidth}
    \includegraphics[width=\textwidth]{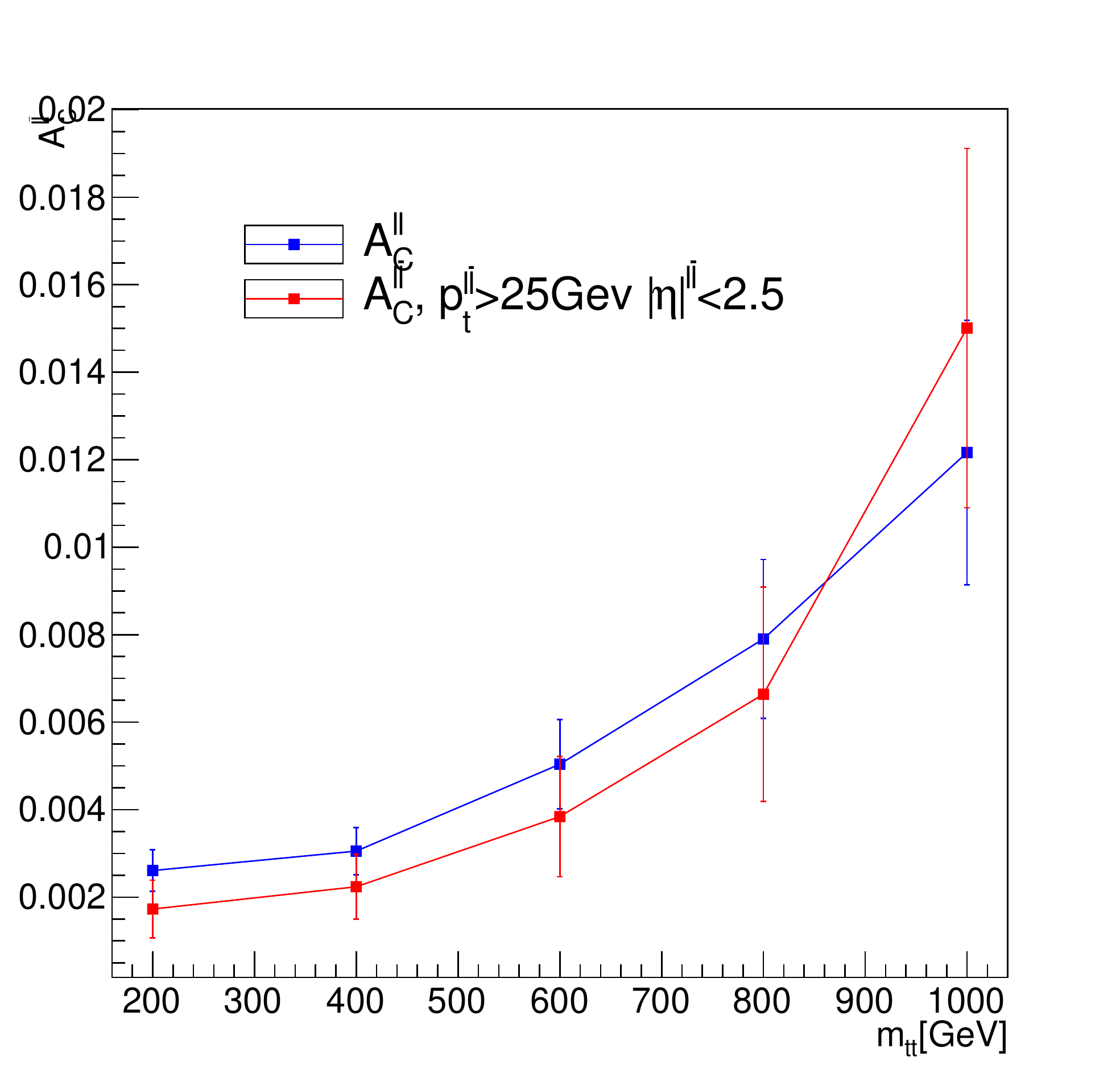}
    \caption{$A_{C}^{l\bar{l}}$ vs $m_{t\bar{t}}$}
    \label{fig:f1}
  \end{subfigure}
  \begin{subfigure}[b]{0.35\textwidth}
    \includegraphics[width=\textwidth]{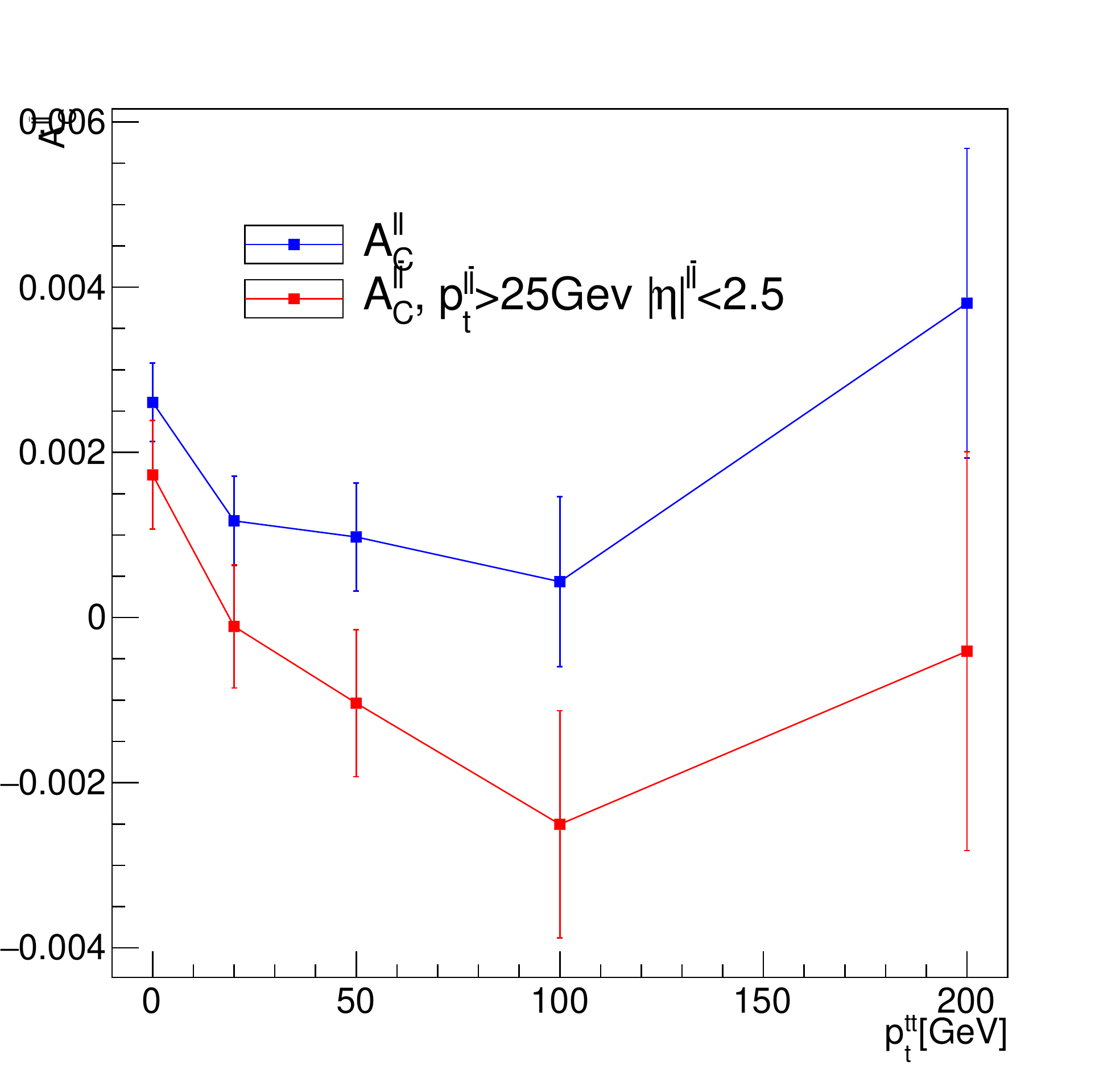}
    \caption{$A_{C}^{l\bar{l}}$ vs $p_{T}^{t\bar{t}}$}
    \label{fig:f2}
  \end{subfigure}
  \begin{subfigure}[b]{0.35\textwidth}
    \includegraphics[width=\textwidth]{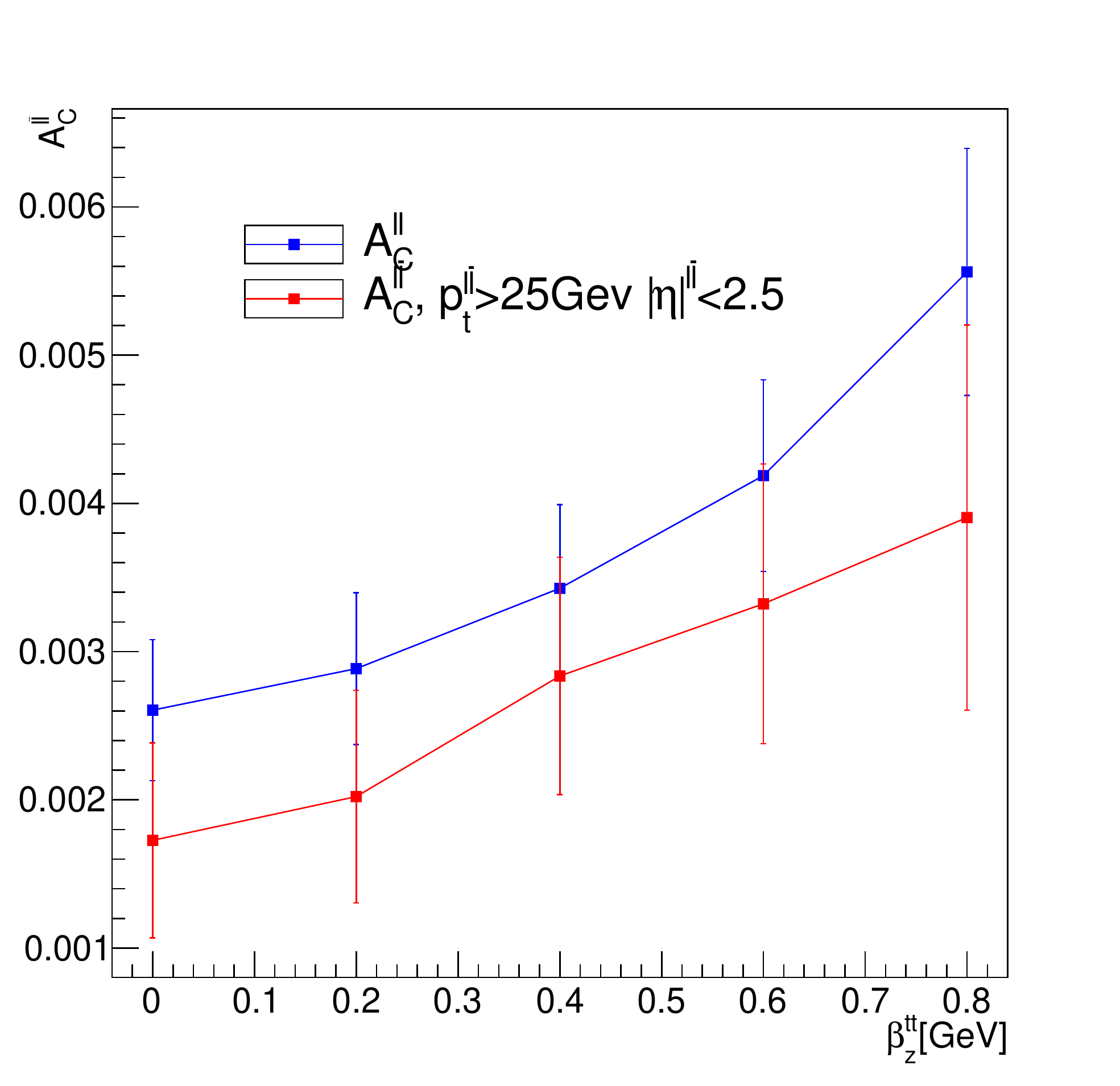}
    \caption{$A_{C}^{l\bar{l}}$ vs $\beta_{z}^{t\bar{t}}$}
    \label{fig:f3}
  \end{subfigure}}
  \caption{Leptonic charge asymmetry as a function of $m_{t\bar{t}}$(left), $p_{T}^{t\bar{t}}$(middle),
$\beta$(right) for lepton pairs from $t\bar{t}$ decay. }
  \label{fig:Integral-asymmetry-ll}
\end{figure}

The asymmetry in $t$$\bar{t}$ is transferred also to the decay products and it can be extracted from the lepton pair only (assuming dilepton decay). The charge asymmetry for lepton pair (leptonic asymmetry) $A_C^{\ell\ell}$ is defined
 the same as for quark pairs  ~\eqref{asym-y},  where $\triangle|y_{l}|=|y_{l^{+}}|-|y_{l^{-}}|$. The lepton asymmetry maintains the same trend as the $t\bar{t}$ asymmetry although the dependence is weaker, see Figure~\ref{fig:Integral-asymmetry-ll}.

\begin{figure}[!h]
\centerline{
  \begin{subfigure}[b]{0.43\textwidth}
    \includegraphics[width=\textwidth]{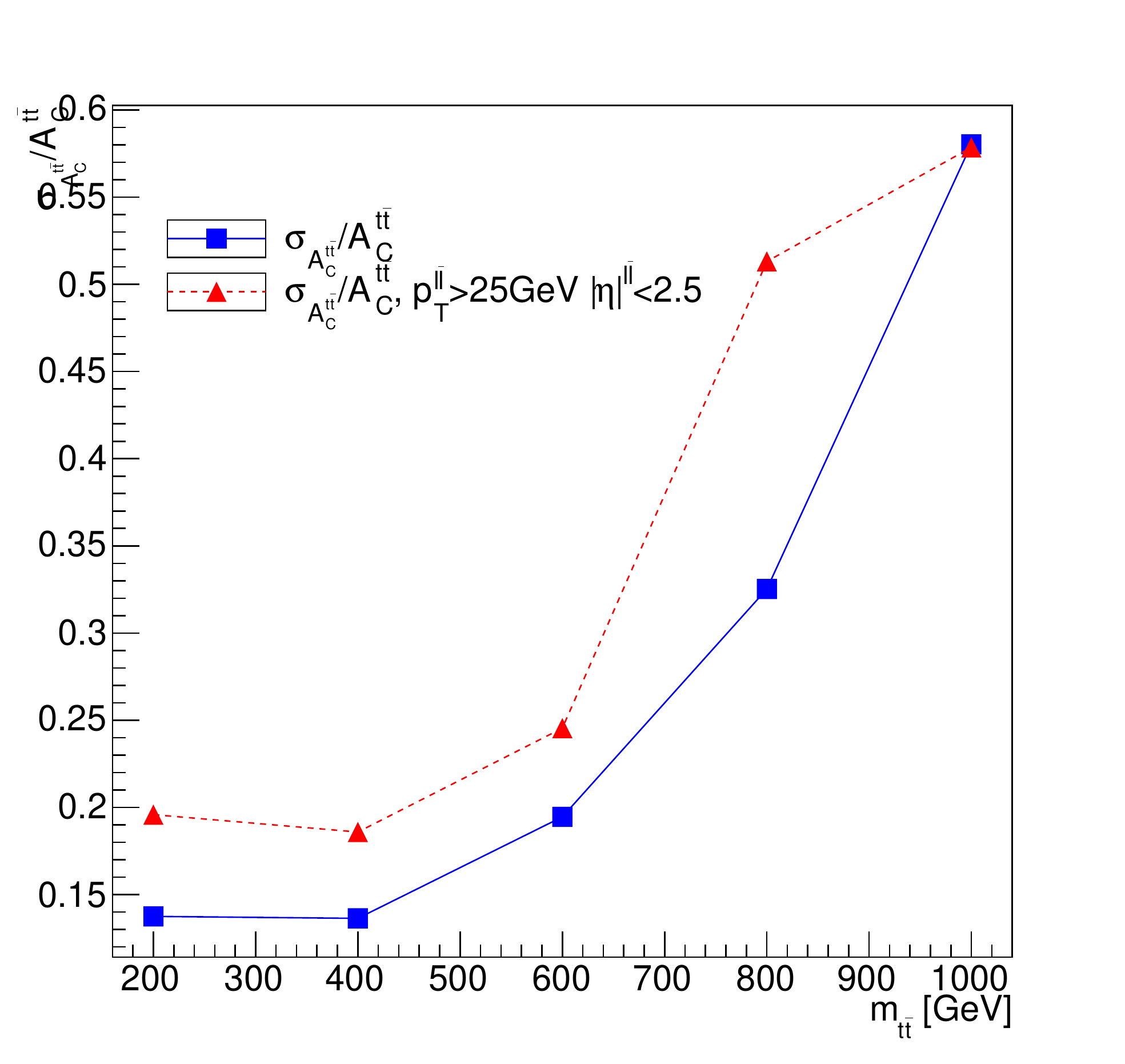}
    \caption{$\sigma$/$A_{C}^{t\bar{t}}$ vs $m_{t\bar{t}}$}
    \label{fig:f1}
  \end{subfigure}
  \begin{subfigure}[b]{0.43\textwidth}
    \includegraphics[width=\textwidth]{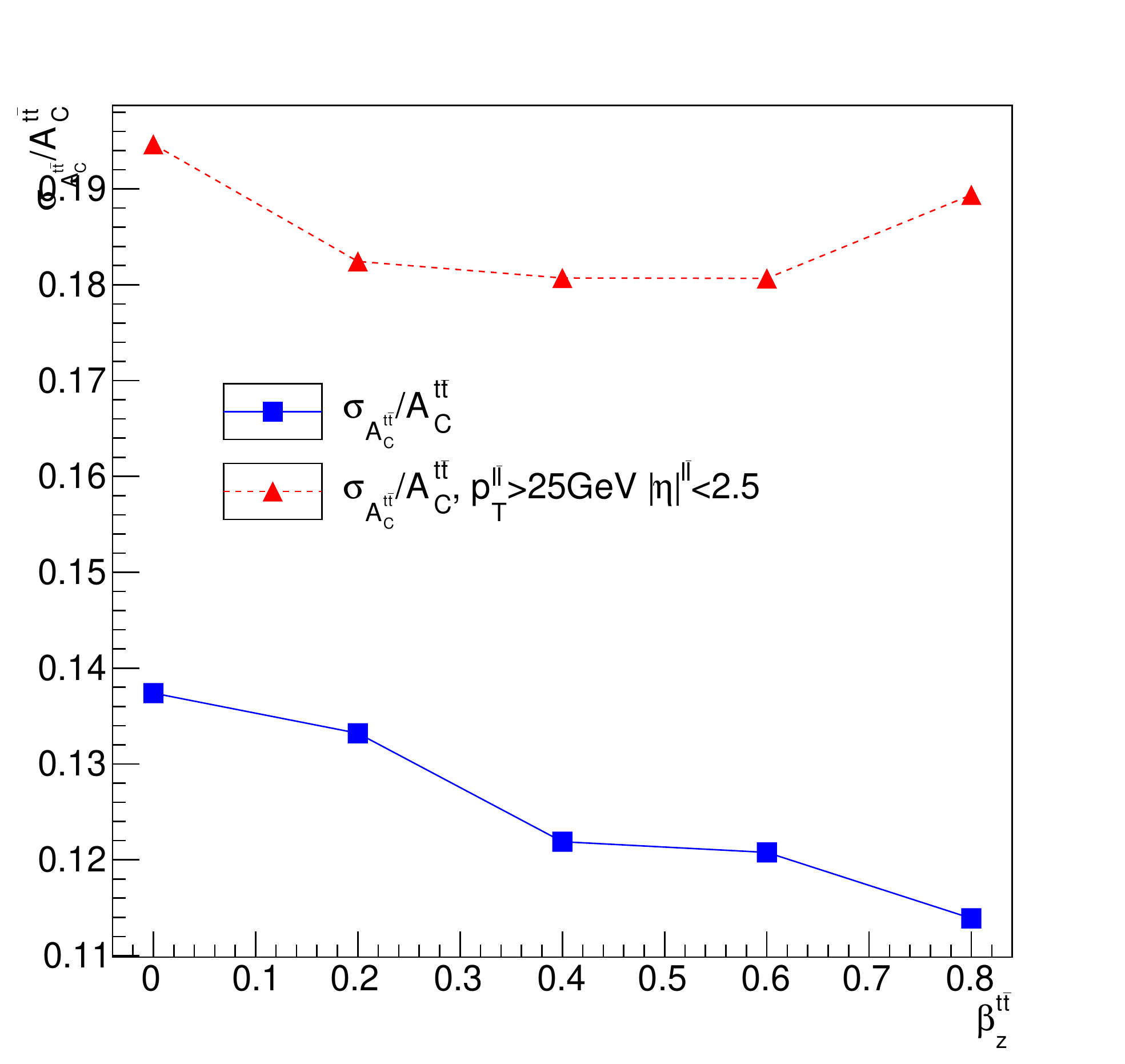}
    \caption{$\sigma$/$A_{C}^{t\bar{t}}$ vs $\beta_{z}^{t\bar{t}}$}
    \label{fig:f2}
  \end{subfigure}}
  \caption{ The estimated relative statistical
  uncertainties for $t\bar{t}$ asymmetry as a function of cut on $m_{t\bar{t}}$
  (a) and $\beta$ (b).}
  \label{fig:allfigs}
\end{figure}

The estimated relative uncertainties for $t\bar{t}$ integral asymmetry corresponding to full expected LHC Run 2 luminosity of
$\L_{i}$=100\,fb$^{-1}$ at $\sqrt{s}$=13\,TeV are shown in Figure~\ref{fig:allfigs}. 
The values of asymmetries were taken from POWHEG simulation while the
statistical uncertainties were rescaled to correspond to $\L_{i}$=100\,fb$^{-1}$. The inclusive $t\bar{t}$ pair production cross section $\sigma(pp \to t\bar{t}+X)$ is assumed to be as 831,76\,pb \cite{cross_section}. The branching ratio of the
dilepton channel (including leptonic $\tau$ decay) was estimated to 0.0676. 
It seems it's best to use the lowest cut (i.e. no
 cut) for $m_{t\bar{t}}$ while to use the largest cut (0.8) for $\beta_{z}^{t\bar{t}}$.
The same conclusion holds also
for the leptonic asymmetry.

It would be useful to perform such studies using the full detector simulation and include the dominant systematic uncertainties in order to optimize $A_{C}$ measurement with real LHC data in Run 2.

\section{Energy and incline asymmetry}

The incline and energy asymmetries \cite{E_incline_asym} were studied for different production  channels. These asymmetries are only defined in $t\bar{t}$+jet system. The results for energy asymmetry are shown in Table~\ref{tab:Results-of-energetic}, the applied jet cuts enhance the asymmetry. Asymmetry in gg channel is consistent with zero as expected.

\begin{table}[h!]
\begin{center}
\caption{Results of energy asymmetry. \label{tab:Results-of-energetic}}
\centerline{%
\begin{tabular}{l|ccc} 
channel & $A^{E}$ & $A^{E}$ ($p_{T}^{j}>25\,GeV$, $|y|<2.5$)\tabularnewline
\hline 
all & $-0.0015\pm0.0003$ & $-0.0032\pm0.0004$\tabularnewline
$q\bar{q}\rightarrow t\bar{t}g$ & $0.002\pm0.001$ & $0.009\pm0.002$\tabularnewline
$qg\rightarrow t\bar{t}q$ & $0.0135\pm0.0007$ & $0.033\pm0.001$\tabularnewline
$\bar{q}g\rightarrow t\bar{t}\bar{q}$ & $0.0148\pm0.0015$ & $0.031\pm0.002$\tabularnewline
$gg\rightarrow t\bar{t}g$ & $0.0006\pm0.0004$ & $0.0006\pm0.0005$\tabularnewline
\hline 
\end{tabular}}
\end{center}
\end{table}

\begin{figure}[h!]
\centerline{\includegraphics[scale=0.38]{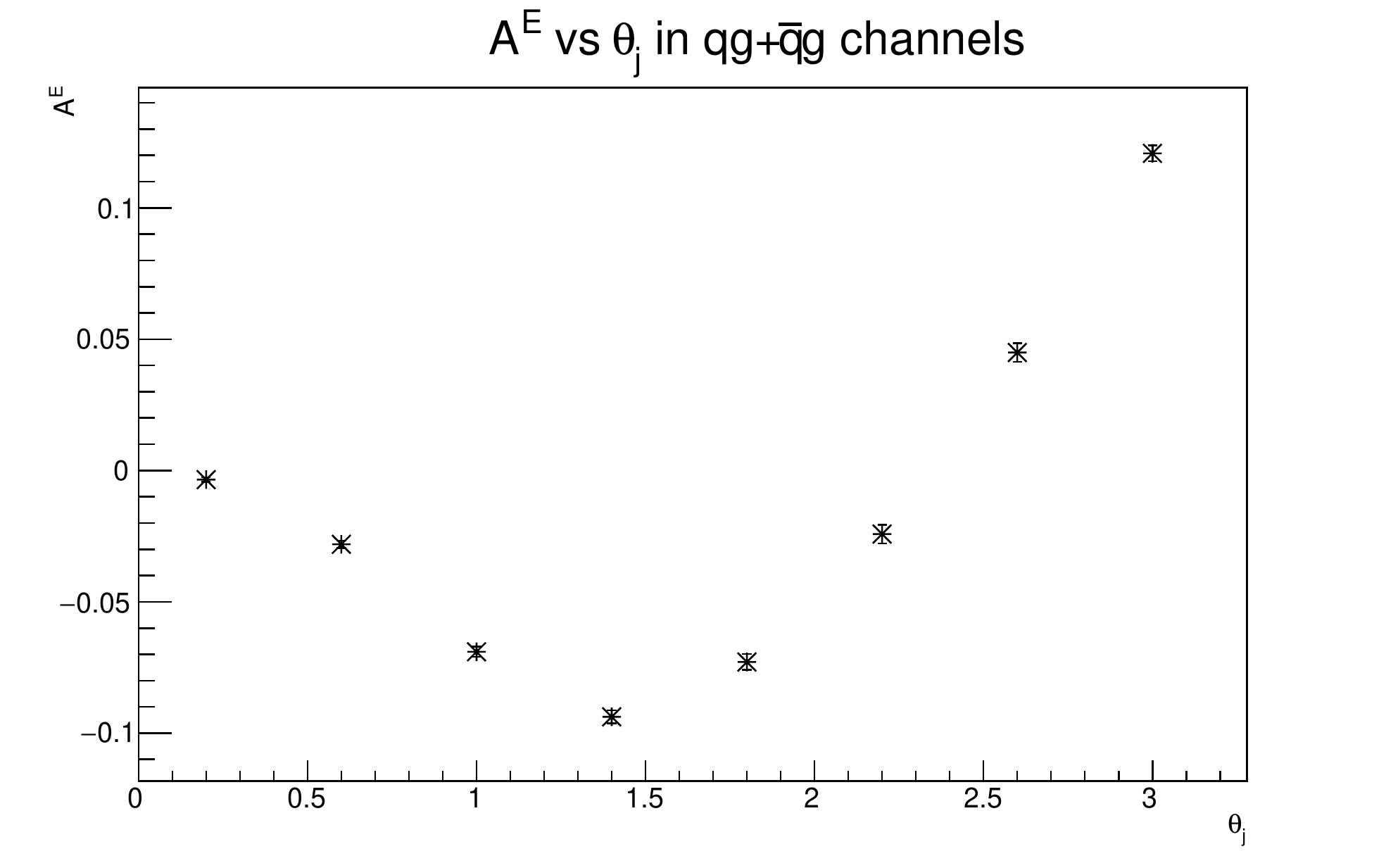}\includegraphics[scale=0.38]{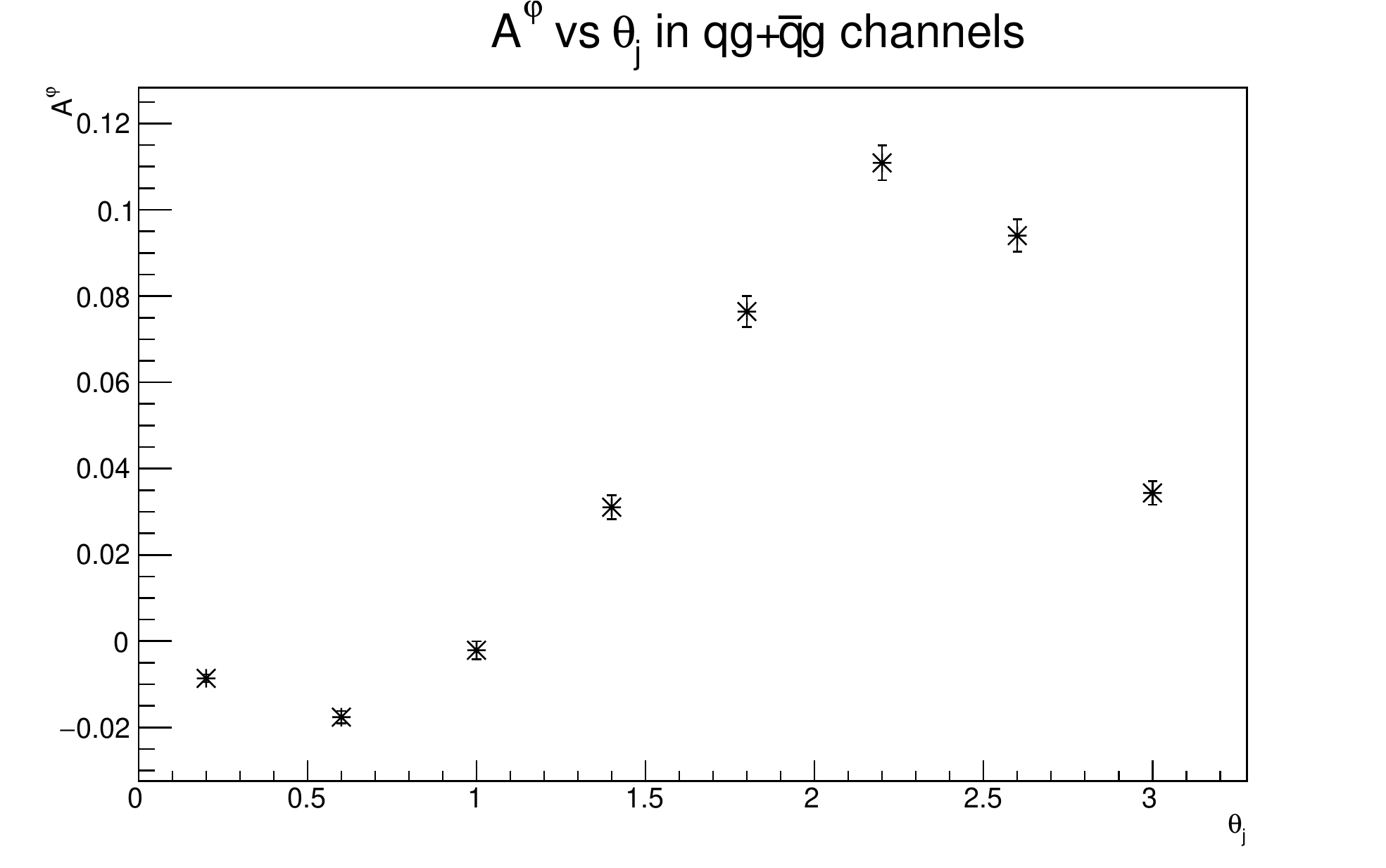}}\caption{Energy $A^{E}$(left) and incline asymmetry $A^{\varphi}$(right) in qg+$\bar{q}$g channels
 as a function
of the jet polar angle $\theta_{j}$.\label{fig:Energy--andinline}}
\end{figure}

For $q$/$\bar{q}$g channel the dependencies of $A^{E}$ and $A^{\varphi}$ on jet polar angle $\theta_{j}$~\cite{E_incline_asym} are shown in Figure~\ref{fig:Energy--andinline}. The minimum in $A^{E}$ dependence as of function of $\theta_{j}$ is around -10\% is in accordance with the analytic results \cite{E_incline_asym}. The minimum and maximum found in $A^{\varphi}$ dependence as of function of $\theta_{j}$ are in qualitative agreement with \cite{E_incline_asym}.
Performing these studies using full detector simulation and including the systematic uncertainties would be helpful to confirm the feasibility of $A^{E}$ and $A^{\varphi}$ measurement at LHC with full Run 2 dataset.

\end{document}

%% file: econfmacros.tex
\def\beq{\begin{equation}}
\def\eeq#1{\label{#1}\end{equation}}
\def\eeqn{\end{equation}}

\def\beqa{\begin{eqnarray}}
\def\eeqa#1{\label{#1}\end{eqnarray}}
\def\eeqan{\end{eqnarray}}

\let\bar=\overbar

\def\L{{\cal L}}

\def\Dslash{\not{\hbox{\kern-4pt $D$}}}
\def\dslash{\not{\hbox{\kern-2pt $\del$}}}

\def\msb{{\bar{\ssstyle M \kern -1pt S}}}

